\documentclass[12pt,preprint]{aastex}
\usepackage{comment}
\usepackage{wrapfig}
\usepackage{float}
\usepackage{sidecap}
\usepackage{natbib}
\usepackage{epstopdf}
\restylefloat{figure}

\shorttitle{VERITAS Observations of Swift J164449.3+573451}
\shortauthors{Aliu et al.}

\begin{document}

\title{VERITAS Observations of the Unusual Extragalactic Transient Swift J164449.3+573451}

\author{
E.~Aliu\altaffilmark{1},
T.~Arlen\altaffilmark{2},
T.~Aune\altaffilmark{3},
M.~Beilicke\altaffilmark{4},
W.~Benbow\altaffilmark{5},
M.~B{\"o}ttcher\altaffilmark{*,6},
A.~Bouvier\altaffilmark{*,3},
S.~M.~Bradbury\altaffilmark{7},
J.~H.~Buckley\altaffilmark{4},
V.~Bugaev\altaffilmark{4},
A.~Cannon\altaffilmark{8},
A.~Cesarini\altaffilmark{9},
L.~Ciupik\altaffilmark{10},
E.~Collins-Hughes\altaffilmark{8},
M.~P.~Connolly\altaffilmark{9},
W.~Cui\altaffilmark{11},
R.~Dickherber\altaffilmark{4},
M.~Errando\altaffilmark{1},
A.~Falcone\altaffilmark{12},
J.~P.~Finley\altaffilmark{11},
L.~Fortson\altaffilmark{13},
A.~Furniss\altaffilmark{3},
N.~Galante\altaffilmark{5},
D.~Gall\altaffilmark{*,14},
G.~H.~Gillanders\altaffilmark{9},
S.~Godambe\altaffilmark{15},
S.~Griffin\altaffilmark{16},
J.~Grube\altaffilmark{10},
G.~Gyuk\altaffilmark{10},
D.~Hanna\altaffilmark{16},
J.~Holder\altaffilmark{17},
H.~Huan\altaffilmark{18},
G.~Hughes\altaffilmark{19},
C.~M.~Hui\altaffilmark{15},
T.~B.~Humensky\altaffilmark{18},
P.~Kaaret\altaffilmark{14},
N.~Karlsson\altaffilmark{13},
M.~Kertzman\altaffilmark{20},
D.~Kieda\altaffilmark{15},
H.~Krawczynski\altaffilmark{4},
F.~Krennrich\altaffilmark{21},
A.~S~Madhavan\altaffilmark{21},
G.~Maier\altaffilmark{19},
P.~Majumdar\altaffilmark{2},
S.~McArthur\altaffilmark{4},
A.~McCann\altaffilmark{16},
P.~Moriarty\altaffilmark{22},
R.~Mukherjee\altaffilmark{1},
R.~A.~Ong\altaffilmark{2},
M.~Orr\altaffilmark{21},
A.~N.~Otte\altaffilmark{3},
N.~Park\altaffilmark{18},
J.~S.~Perkins\altaffilmark{5},
A.~Pichel\altaffilmark{23},
M.~Pohl\altaffilmark{19, 24},
H.~Prokoph\altaffilmark{19},
J.~Quinn\altaffilmark{8},
K.~Ragan\altaffilmark{16},
L.~C.~Reyes\altaffilmark{18},
P.~T.~Reynolds\altaffilmark{25},
E.~Roache\altaffilmark{5},
H.~J.~Rose\altaffilmark{7},
J.~Ruppel\altaffilmark{24},
D.~B.~Saxon\altaffilmark{17},
M.~Schroedter\altaffilmark{5},
G.~H.~Sembroski\altaffilmark{11},
C.~Skole\altaffilmark{19},
A.~W.~Smith\altaffilmark{26},
D.~Staszak\altaffilmark{16},
G.~Te\v{s}i\'{c}\altaffilmark{16},
M.~Theiling\altaffilmark{5},
S.~Thibadeau\altaffilmark{4},
K.~Tsurusaki\altaffilmark{14},
J.~Tyler\altaffilmark{16},
A.~Varlotta\altaffilmark{11},
S.~Vincent\altaffilmark{15},
M.~Vivier\altaffilmark{17},
S.~P.~Wakely\altaffilmark{18},
J.~E.~Ward\altaffilmark{8},
A.~Weinstein\altaffilmark{2},
T.~Weisgarber\altaffilmark{18},
D.~A.~Williams\altaffilmark{3}
}

\altaffiltext{*}{Corresponding authors: M. B{\"o}ttcher, boettchm@ohio.edu; A. Bouvier, apbouvie@ucsc.edu; D. Gall, daniel-d-gall@uiowa.edu.}
\altaffiltext{1}{Department of Physics and Astronomy, Barnard College, Columbia University, NY 10027, USA}
\altaffiltext{2}{Department of Physics and Astronomy, University of California, Los Angeles, CA 90095, USA}
\altaffiltext{3}{Santa Cruz Institute for Particle Physics and Department of Physics, University of California, Santa Cruz, CA 95064, USA}
\altaffiltext{4}{Department of Physics, Washington University, St. Louis, MO 63130, USA}
\altaffiltext{5}{Fred Lawrence Whipple Observatory, Harvard-Smithsonian Center for Astrophysics, Amado, AZ 85645, USA}
\altaffiltext{6}{Astrophysical Institute, Department of Physics and Astronomy, Ohio University, Athens, OH 45701}
\altaffiltext{7}{School of Physics and Astronomy, University of Leeds, Leeds, LS2 9JT, UK}
\altaffiltext{8}{School of Physics, University College Dublin, Belfield, Dublin 4, Ireland}
\altaffiltext{9}{School of Physics, National University of Ireland Galway, University Road, Galway, Ireland}
\altaffiltext{10}{Astronomy Department, Adler Planetarium and Astronomy Museum, Chicago, IL 60605, USA}
\altaffiltext{11}{Department of Physics, Purdue University, West Lafayette, IN 47907, USA }
\altaffiltext{12}{Department of Astronomy and Astrophysics, 525 Davey Lab, Pennsylvania State University, University Park, PA 16802, USA}
\altaffiltext{13}{School of Physics and Astronomy, University of Minnesota, Minneapolis, MN 55455, USA}
\altaffiltext{14}{Department of Physics and Astronomy, University of Iowa, Van Allen Hall, Iowa City, IA 52242, USA}
\altaffiltext{15}{Department of Physics and Astronomy, University of Utah, Salt Lake City, UT 84112, USA}
\altaffiltext{16}{Physics Department, McGill University, Montreal, QC H3A 2T8, Canada}
\altaffiltext{17}{Department of Physics and Astronomy and the Bartol Research Institute, University of Delaware, Newark, DE 19716, USA}
\altaffiltext{18}{Enrico Fermi Institute, University of Chicago, Chicago, IL 60637, USA}
\altaffiltext{19}{DESY, Platanenallee 6, 15738 Zeuthen, Germany}
\altaffiltext{20}{Department of Physics and Astronomy, DePauw University, Greencastle, IN 46135-0037, USA}
\altaffiltext{21}{Department of Physics and Astronomy, Iowa State University, Ames, IA 50011, USA}
\altaffiltext{22}{Department of Life and Physical Sciences, Galway-Mayo Institute of Technology, Dublin Road, Galway, Ireland}
\altaffiltext{23}{Instituto de Astronomia y Fisica del Espacio, Casilla de Correo 67 - Sucursal 28, (C1428ZAA) Ciudad Aut—noma de Buenos Aires, Argentina}
\altaffiltext{24}{Institut f\"ur Physik und Astronomie, Universit\"at Potsdam, 14476 Potsdam-Golm,Germany}
\altaffiltext{25}{Department of Applied Physics and Instrumentation, Cork Institute of Technology, Bishopstown, Cork, Ireland}
\altaffiltext{26}{Argonne National Laboratory, 9700 S. Cass Avenue, Argonne, IL 60439, USA}

\begin{abstract}

We report on very-high-energy ($>$100~GeV) gamma-ray observations of Swift J164449.3+573451, an unusual transient object first detected by the {\it Swift} Observatory and later detected by multiple radio, optical and X-ray observatories. A total exposure of 28 hours was obtained on Swift J164449.3+573451 with VERITAS during 2011 March 28 -- April 15.  We do not detect the source and place a differential upper limit on the emission at 500~GeV during these observations of $1.4 \times 10^{-12}$ erg cm$^{-2}$ s$^{-1}$ ($99\%$ confidence level).  We also present time-resolved upper limits and use a flux limit
averaged over the X-ray flaring period to constrain various emission scenarios that can accommodate both the radio-through-X-ray emission detected from the source and the lack of detection by VERITAS.

\end{abstract}

\section{Introduction}\label{sec:Introduction}

Swift J164449.3+573451 (hereafter, Sw\,J1644+57) was first detected by the {\it Swift} Burst Alert Telescope (BAT) on 2011 March 28 at 12:57:45 UT.  The {\it Swift} spacecraft slewed to the location of the source and began observations with the X-ray Telescope (XRT) and the UV/Optical Telescope (UVOT).  These observations located a bright, uncatalogued X-ray source but did not identify an optical afterglow typical of gamma-ray bursts (GRBs) \citep{firstdet}.  Less than one hour later, the BAT triggered a second time on Sw\,J1644+57,  which
ruled out a GRB origin and gave the first sign of the unusual nature of the source \citep{swiftsfxt}. This prompted multiwavelength follow-up observations at a number of observatories.

These follow-up observations identified an optical source consistent with the position of Sw\,J1644+57 \citep{ptf,not}.  Measurements obtained with the Gemini Observatory show an infrared (IR) source with a transient component at a location consistent with that of Sw\,J1644+57 and provide a redshift of $z=0.3534$ from H$_{\beta}$ and OIII emission lines \citep{gemini, geminib}.  
Hubble Space Telescope observations show a nearly point-like IR source consistent with the location of Sw\,J1644+57 and, in an optical exposure, a resolved compact galaxy whose nucleus is consistent with the position of the IR point source \citep{hubble}.  Radio observations with the Enhanced Very Large Array find an unresolved, variable radio source at a position consistent with the nucleus of the host galaxy suggested by the Hubble observations \citep{evla1,evla2, astrometry}.  Temporal analysis of the {\it Swift} light curve, combined with the implied peak luminosity at a distance of $z \sim 0.35$, provides evidence that the observed emission from Sw\,J1644+57 is likely beamed \citep{beaming, swiftpaper, disruption}.

Because X-ray and very-high-energy (VHE; E$>$100 GeV) gamma-ray emission are frequently correlated in other beamed sources, such as blazars \citep{boettcher10}, it is reasonable to expect VHE emission from Sw\,J1644+57, depending on the parameters of the emission region and the surrounding environment.  Here, we discuss deep VHE observations of Sw\,J1644+57 with the Very Energetic Radiation Imaging Telescope Array System (VERITAS) and the implications of our results for some possible emission scenarios for this unusual object.

\section{Observations}
\label{sec:Analysis}

VERITAS is an array of four imaging atmospheric-Cherenkov telescopes (IACTs) located at the Fred Lawrence Whipple Observatory in southern Arizona at an altitude of 1280m above sea level \citep{veritasstatus}.  Imaging cameras, consisting of 499 photomutiplier tubes located in the focal plane of each telescope, detect Cherenkov light emitted by extensive air showers initiated in the upper atmosphere by gamma rays and cosmic rays. VERITAS has a field of view of 3.5$^{\circ}$ and is sensitive in the range of 100 GeV -- 30 TeV.  The telescopes typically operate in ``wobble" mode, where the location of the target is offset from the center of the field of view by $0.5^{\circ}$, allowing for simultaneous background measurements \citep{Fomin}.  The offset direction alternates between north, south, east and west for each data segment (typically lasting 20 minutes) to reduce systematic errors in the background estimation.

On 2011 March 29 at 10:27 UT, approximately 22.5 hours after the first BAT trigger, VERITAS started observing Sw\,J1644+57. Subsequent daily observations with an average exposure of $\sim$2 hours/night were taken when weather conditions were favorable, continuing through 2011 April 15, after which observations were not possible because of the near-full Moon (exceeding $\sim$97\% illumination).  
Zenith angles for our observations ranged from $25^{\circ}$ to $40^{\circ}$.  Due to temporary hardware issues, approximately 15\% of the data were taken with an array of three telescopes. In total, VERITAS accumulated $\sim$28 hours of exposure on this source, of which $\sim$3.5 hours were taken within one day of the particularly intense flaring events observed in X-rays during 2011 March 28--31 \citep{swiftdetail}.

For this analysis, about 90\% of the data ($\sim$25 hours) pass the quality selection criteria, with selection based primarily on weather conditions and trigger-rate stability.  The selected data are processed through the standard VERITAS analysis package \citep{vegas}.
Our cosmic-ray rejection procedure is based on applying selection criteria on standard image parameters \citep{hillas}:  the size of the telescope images, the mean scaled width and mean scaled length parameters \citep{Krawczynski(2006)}, the height of maximum Cherenkov emission and the angular distance from the putative source position to the reconstructed arrival direction of the shower ($\theta$). The standard selection criteria (see Table \ref{VERITAScuts}) were optimized using Monte Carlo simulations and real data from the Crab Nebula and the blazar PG 1553+113 .

The remaining background is estimated using the ``reflected-region" method described in \citet{reflectedreg}.  The radii of the circular on- and off-source regions are $0.1^{\circ}$.  Statistical significances are computed using a modified version of Equation 17 from \citet{LiMa} to allow for varying number of off-source regions due to the bright ($V=4.849$) nearby star HR 6237 \citep{mixedalpha}.

\section{Results} \label{sec:Results}

Significant VHE gamma-ray emission is not detected from the direction of Sw\,J1644+57 in the entire data set nor in subsets of the data (see Table \ref{results}).  In order to look specifically for VHE emission contemporaneous with the intense X-ray flaring, the first subset consists of data that were taken within one day before or after periods where the XRT count rate exceeded  20 s$^{-1}$. This subset is denoted the ``flaring" period and comprises the first three nights of observations.  However, it is worth noting that VERITAS exposures during this ``flaring'' period fell between X-ray flares observed by {\it Swift}.  Therefore, VERITAS observations were simultaneous with relatively low X-ray flux states during that period, {characterized by
an X-ray flux of $\nu F_{\nu} \sim 10^{-10}$~erg~cm$^{-2}$~s$^{-1}$,
about two orders of magnitude lower than the major flares.} The second subset, denoted the ``low" period, comprises the remainder of the data.

%In this period, a significance of $-0.3$ standard deviations was measured.  

Following the lack of signal in the data, we derive 99\% confidence level upper limits over various time intervals (see Table \ref{results}):

\begin{itemize}
\item Total:  2011 March 29 -- April 15
\item Flaring: 2011 March 29 -- March 31
\item Low: 2011 April 1 -- April 15
\item Daily (by UT date, when observations available)
\end{itemize}

\begin{figure}[t]
\centering
\includegraphics[height=5in, angle=90]{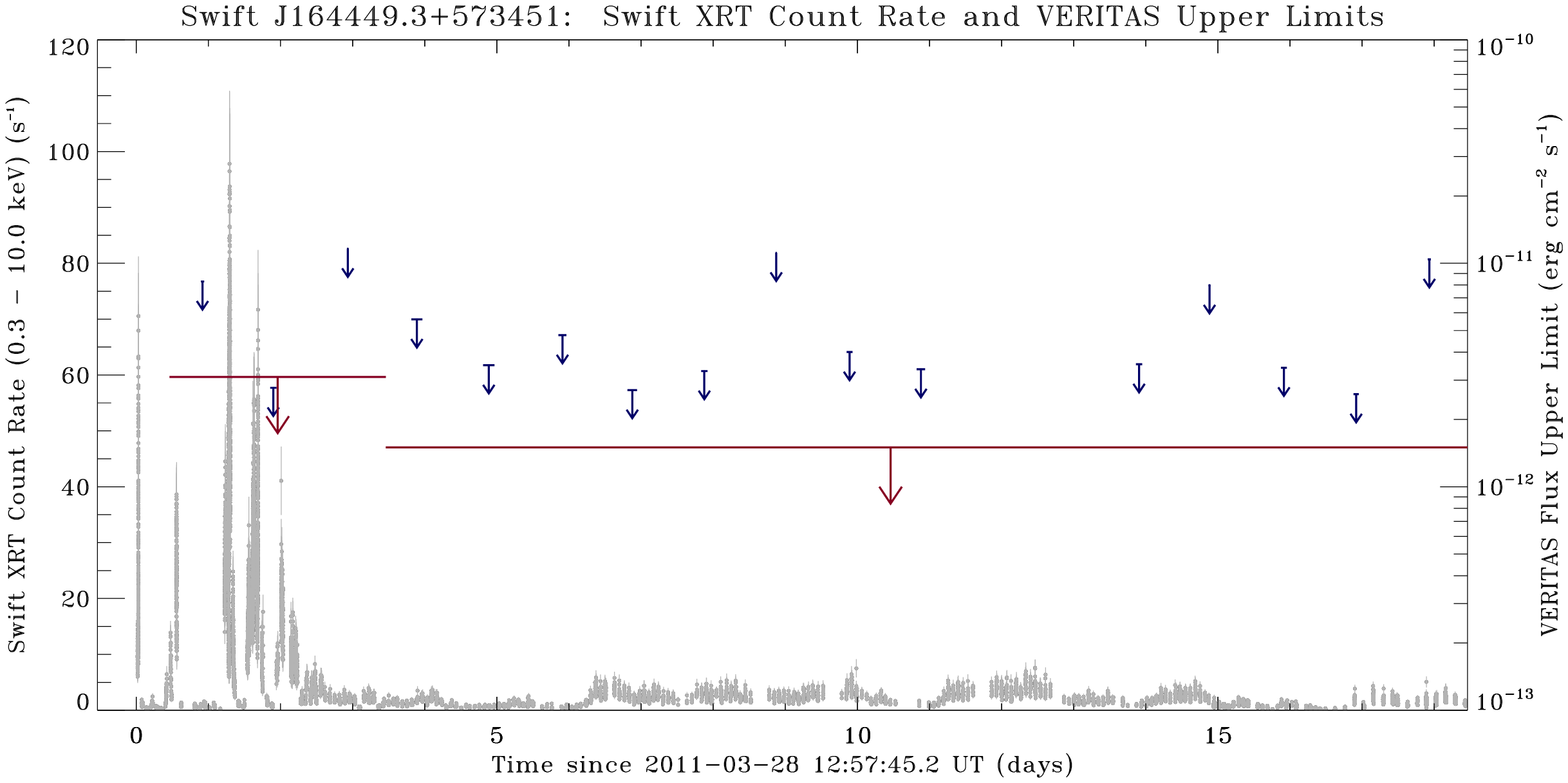}
\caption{Flaring, low-state and daily upper limits are shown superimposed on the {\it Swift} XRT light curve \citep{xrtcurve}. The extent of the daily upper limit horizontal bars represents the approximate time interval during which the VERITAS observations were taken.}
\label{swiftlc}
\end{figure}

The procedure described by \citet{Rolke} is chosen for the upper-limit computation with the assumption of a Gaussian-distributed background.  The total, flaring and low-flux upper limits on {\it E*F(E)} at 99\% c.l. are: $1.4 \times 10^{-12}$, $3.1 \times 10^{-12}$ and $1.5 \times 10^{-12}$ erg\,cm$^{-2}$\,s$^{-1}$, respectively, where {\it F(E)} is the energy flux.  The limits are calculated at 500 GeV assuming any emission follows a power-law spectrum with a photon index of -3.0.  The decorrelation energy (500 GeV) is used to reduce the sensitivity of the limits to the choice of photon index; this energy is higher than the energy threshold of the observations ($\sim$290 GeV).  The flaring and low-state upper limits along with the daily upper limits are presented in Figure \ref{swiftlc}, superimposed on the {\it Swift} XRT light curve \citep{xrtcurve} for comparison.

\section{Discussion}
\label{interpretation}

In this section, we provide some generic parameter constraints that can
be derived from the observed X-ray properties of Sw\,J1644+57, along with 
the non-detection by {\it Fermi}/LAT \citep{fermi} and VERITAS.  The X-ray flux varied on timescales of $t_{\rm var}=100$~seconds
\citep{swiftpaper}, with a peak energy flux of 
$F_X\sim10^{-8}$~erg~cm$^{-2}$~s$^{-1}$,  corresponding to a peak luminosity 
of $L_{\rm pk} \sim 4.3 \times 10^{48}$~erg~s$^{-1}$ if the emission were isotropic. 
In order to illustrate the dependence of the following estimates
on the variability time scale, we parameterize $t_{\rm var} \equiv 
100\,t_{{\rm var,}2}$~seconds.
The Eddington limit implies a central engine mass of $M>3.4\times10^{10}\,M_{\odot}$, assuming 
unbeamed emission. 
Assuming the emission-region size is not smaller than the Schwarzschild radius of the central engine, the observed variability 
implies $M<10^7M_{\odot}$. The two mass estimates can be reconciled by allowing
for anisotropic and/or beamed emission, plausibly involving relativistic motion.
Relativistic motion will result in Doppler boosting of the luminosity by a factor $D^4$, 
along with variability time contraction by a factor $D^{-1}$, where $D=\left(
\Gamma[1-\beta_{\Gamma}\cos\theta] \right)^{-1}$ is the Doppler factor,
$\Gamma=(1-\beta_{\Gamma}^2)^{-1/2}$ is the bulk Lorentz factor of the 
emission region, $\beta_{\Gamma}c$ is its velocity, and $\theta$ is the angle between the direction of motion and the line of sight.  Reconciling the mass estimates above requires Doppler 
boosting by at least a factor $D>5.4$.  

\citet{disruption} and \citet{swiftpaper} have argued that this event arises 
from the activation of a beamed jet and have hypothesized that this may be 
the result of tidal disruption of a star by a $\sim 10^6$--$10^7 
\, M_{\odot}$ black hole. Both synchrotron- \citep{swiftpaper} and Compton-dominated
 \citep{disruption} origins have been proposed for the X-ray emission.
\cite{swiftpaper} propose a Poynting-flux-dominated scenario, in which the 
X-ray emission is produced by synchrotron emission from relativistic electrons.
\cite{disruption} interpret the lack of variability of the radio -- IR 
emission as evidence that the radio -- IR emission is produced in a more extended 
region than the X-rays. They suggest inverse-Compton scattering of external 
radiation as the mechanism producing the high-energy radiation. 

In the following discussion, we present some general considerations to 
constrain the parameters of the X-ray emission region, including 
constraints placed by the VHE upper limits. We consider both 
synchrotron and inverse-Compton as possible emission mechanisms.

\subsection{\label{synchrotron}Synchrotron Origin}

We first consider a scenario in which the X-ray emission is synchrotron
emission by relativistic electrons in a tangled magnetic field $B$. The electron 
Lorentz factor at which the non-thermal electron 
distribution has its peak radiative output is $\gamma_p$, and the
electron density at that energy is $n_p\equiv n_e(\gamma_p)$. The 
observed spectral variability suggests
that the peak frequency might vary substantially during
the various outbursts. For the following estimates, we scale the peak 
frequency as $\nu_{\rm pk} \equiv 10^{19}\,\nu_{p, 19}$~Hz and we base
our estimates on the typical X-ray flux observed during VERITAS observations in the flaring state $\nu F_{\nu}^{\rm sy} \equiv 10^{-10} \, f_{-10}$~erg~cm$^{-2}$~s$^{-1}$, corresponding to $\nu L_{\nu} 
= 4.3\times10^{46}$~erg~s$^{-1}$. We further assume that the 
variability timescale provides an estimate of the emission-region size, 
$R_B = c \, t_{\rm var} \, D / (1 + z)$. The observables can then be 
related to the emission-region parameters through \citep{rybicki}:

\begin{equation}
\nu_{\rm pk} = 4.2 \times 10^6 \, \gamma_p^2 \, {\left(\frac{B}{\rm 1\ G}\right)} 
\, {D \over 1 + z}
\; {\rm Hz} \, ,
\label{nupk}
\end{equation}

\begin{equation}
\nu L_{\nu} = {2 \over 9} \, c \, \sigma_T \, B^2 \, \gamma_p^2
\, n_p \, \left( {c \, t_{\rm var} \over 1 + z} \right)^3 \, D^7 \, .
\label{Lpk}
\end{equation}

A further constraint is derived from the condition that the
synchrotron cooling timescale of electrons of energy $\gamma_p$ 
should be of the order of the observed variability timescale. 
This corresponds to the assumption that the entire energy
transferred to radiation throughout the duration of the flare
is contained in the particle population at the onset of the 
flare. We define such a scenario as a {\it particle-dominated
scenario}. 
\cite{swiftpaper} have shown that an alternative, Poynting-flux-dominated scenario with synchrotron cooling timescales of the order 
of $t_{\rm sy}\lesssim0.1$~s can explain the spectral energy distribution (SED) of Sw\,J1644+57.
Such a scenario requires continuous {\it in situ} re-acceleration of 
electrons to maintain a low-energy cut-off in the electron distribution, 
which is needed in order to reproduce the observed hard optical -- X-ray 
spectral slope. 

Assuming that the electron cooling timescale and the light-crossing 
timescale across the source are of the same order, we estimate

\begin{equation}
t_{\rm var} \sim t_{\rm sy} \, {1 + z \over D} = {6 \, \pi \, m_e c^2 \over c \, \sigma_T \, B^2 \, \gamma_p} \, {1 + z \over D} \, .
\label{tcool}
\end{equation}

\noindent Parameterizing the Doppler factor in terms of $D_1 \equiv D/10$, we
solve Equations \ref{nupk} -- \ref{tcool} to find

\begin{eqnarray}
B &=& 1.5 \, D_1^{-1/3} \, \nu_{p, 19}^{-1/3} \, t_{{\rm var,}2}^{-2/3} \; {\rm G} \, ,
\label{B}
\end{eqnarray}

\begin{eqnarray}
\gamma_p &=& 4.4 \times 10^5 \, D_1^{-1/3} \, t_{{\rm var,}2}^{1/3} \, \nu_{p, 19}^{2/3} \, ,
\label{gamma}
\end{eqnarray}

\begin{eqnarray}
n_p &=& 6.7 \times 10^4 \, D_1^{-17/3} \, \nu_{p, 19}^{-2/3} \, t_{{\rm var,}2}^{-7/3}
\, f_{-10} \; {\rm cm}^{-3} \, .
\label{n_p}
\end{eqnarray}

\noindent These parameters correspond to a Thomson depth from electrons near the peak, $\tau_T$, of

\begin{equation}
\tau_T = n_e \, \sigma_T \, R_B = 10^{-6} \, D_1^{-14/3} \, 
\nu_{p, 19}^{-2/3} \, t_{{\rm var,}2}^{-4/3} \, f_{-10} \, .
\label{tauT}
\end{equation}

\noindent The expected synchrotron radiation energy density in the
co-moving frame, $u'_{\rm sy}$, is 

\begin{equation}
u'_{\rm sy} \sim {16 \over 9} \, \tau_T \, \gamma_p^2 \, u'_B \, ,
\label{usy}
\end{equation}
where $u'_B = B^2 / (8\pi)$ is the co-moving magnetic-field energy density. 
Electrons with energy corresponding to $\gamma_p$ will encounter synchrotron photons at 
frequency $\nu_{\rm pk}$ (observer's frame) in the extreme Klein-Nishina limit, $\gamma_p \, {\epsilon'}_{\rm pk} 
\sim 3.5 \times 10^3 \, D_1^{-4/3} \, t_{{\rm var,}2}^{1/3} \, \nu_{p, 19}^{5/3}$ where ${\epsilon'}_{\rm pk} = h \nu_{\rm pk} / 
(D \, m_e c^2)$. Therefore, the 
inverse-Compton output from electrons at $\gamma_p$ is strongly suppressed by
a factor $f_{\rm KN} = \frac{3}{8} \frac{ln(2 \epsilon' \gamma) + 1/2}{\epsilon'\gamma} 
\sim 7.6 \times 10^{-4} \, D_1^{4/3} \, t_{{\rm var,}2}^{-1/3} \, \nu_{p, 19}^{-5/3}$.
Hence, the ratio of synchrotron self-Compton (SSC) to synchrotron luminosities 
is expected to be

\begin{equation}
\left( {L_{\rm SSC} \over L_{\rm sy} } \right) = {u'_{\rm sy} \over u'_B} \times
f_{\rm KN} \sim 2.6 \times 10^2 \, D_1^{-4} \, t_{{\rm var,}2}^{-1} \, \nu_{p, 19}^{-1} \, f_{-10} \, ,
\label{Lratio_KN}
\end{equation}
which leads to an estimated intrinsic VHE flux of

\begin{equation}
\nu F_{\nu}^{\rm SSC, int} = \left( {L_{\rm SSC} \over L_{\rm sy}}
\right) \, \nu F_{\nu}^{\rm sy} 
\sim 2.6 \times 10^{-8} \, D_1^{-4} \, t_{{\rm var,}2}^{-1} \, 
\nu_{p, 19}^{-1} \, f_{-10}^2 \; {\rm erg} \; {\rm cm}^{-2} \; 
{\rm s}^{-1} \, .
\label{nuFnuSSC}
\end{equation}
with an emission peak around $D \gamma_p m_e c^2/(1 + z) \sim 1.7 D_1^{2/3} \,
t_{{\rm var,}2}^{1/3} \, \nu_{p, 19}^{2/3}$~TeV.
We point out that if $L_{\rm SSC}/L_{\rm sy} > 1$, the effective electron cooling timescale will actually be shorter than the synchrotron cooling timescale by a factor 
$\sim L_{\rm sy}/L_{\rm SSC}$, in which case our initial assumption $t_{\rm var} \sim t_{\rm sy} \, {1 + z \over D}$ breaks down.

At VHE photon energies, the effect of $\gamma\gamma$ absorption internal to the emission region may become substantial 
\citep{swiftpaper}. Photons of co-moving photon energy 
${\epsilon'}_{\gamma} \equiv h {\nu'}_{\gamma} / (m_e c^2)
\equiv 10^5 \, \epsilon_6 \, D_1^{-1}$ are most efficiently absorbed by 
target photons of energy ${\epsilon'}_T \sim 2/{\epsilon'}_{\gamma}$, 
corresponding to an observed target photon frequency of 
$\nu_T \sim 10^{16} \, D_1^2 \, \epsilon_6^{-1}$ Hz, i.e., UV photons.
We note that the UV flux appears to be very strongly absorbed by gas and dust local
to the host galaxy, and the intrinsic UV flux may well be several orders of magnitude
higher than the UVOT upper limits of $\lesssim 10^{-14}$~erg~cm$^{-2}$~s$^{-1}$. Here we parameterize the intrinsic flux of photons at those frequencies as $\nu F_{\nu} (\epsilon_T) 
\equiv 10^{-11} \, f_{UV, -11}$~erg~cm$^{-2}$~s$^{-1}$ since the X-ray to optical flux extrapolations from the SED, as well as other evidence about the extinction, show that $f_{UV}$ is probably around $0.5 \times 10^{-11} \; {\rm erg} \; {\rm cm}^{-2} \; {\rm s}^{-1}$ \citep{swiftpaper}.
Based on a $\delta$-function approximation to the $\gamma\gamma$ absorption cross-section, 
the optical depth for $\gamma\gamma$ absorption is estimated as

\begin{equation}
\tau_{\gamma\gamma}^{\rm int} \sim {4 \over 3} \, {\sigma_T \, d_L^2 \, \nu F_{\nu} (\epsilon_T)
\, (1 + z) \over \epsilon'_T \, D^5 \, m_e c^2 \, c^2 \, t_{\rm var} }
\sim 5.5 \times 10^3 \, f_{UV, -11} \, \epsilon_6 \, D_1^{-6} \, t_{{\rm var,}2}^{-1} \, .
\label{taugg}
\end{equation}

For internal absorption, the suppression of the flux is given 
by $F_{\rm abs, internal} = F_{\rm internal} (1 - 
e^{-\tau_{\gamma\gamma}^{\rm int}}) / \tau_{\gamma\gamma}^{\rm int} 
\approx F_{\rm int} / \tau_{\gamma\gamma}^{\rm int}$ for 
$\tau_{\gamma\gamma}^{\rm int} \gg 1$. Consequently, after correction
for extinction by the extragalactic background light (EBL), which 
amounts to a factor of $e^{-\tau_{\gamma\gamma}^{\rm EBL}} \sim 1/13$ 
at 500~GeV, using the \cite{finke10} EBL model, the particle-dominated 
synchrotron scenario predicts a VHE $\gamma$-ray flux of 
$\nu F_{\nu}^{\rm SSC, int} \, e^{-\tau_{\gamma\gamma}^{\rm EBL}}
/ \tau_{\gamma\gamma}^{\rm int} \sim 3.6 \times 
10^{-13}$~erg~cm$^{-2}$~s$^{-1}$, which is slightly below the 
VERITAS upper limits.

However, such an emission model would require the following equipartition ratio between the co-moving energy densities
in the magnetic field, $u'_B$, and the relativistic electrons,
$u'_e$, based on the parameters of Equations \ref{B} - \ref{n_p}:

\begin{equation}
\epsilon_B \equiv u'_B / u'_e = 3.7 \times 10^{-6}  D_1^{16/3}  \nu_{p, 19}^{-2/3}  t_{{\rm var,}2}^{2/3} f_{-10}^{-1} \, .
\label{equipartition}
\end{equation}

\noindent Therefore, the particle-dominated synchrotron scenario, though possible, is disfavored as it requires an unusually large Doppler factor ($\gtrsim 100$) to allow for equipartition to occur in the jet.

\subsection{\label{IC}External Inverse-Compton Origin}

The X-rays may also be produced by inverse-Compton scattering 
of low-energy radiation. We scale the peak frequency of the soft 
target photons for Compton scattering as $\nu_s \equiv
10^{13} \, \nu_{s,13}$~Hz. Assuming that the external radiation field
is approximately isotropic in the rest frame of the host galaxy, 
the observed Compton peak frequency is $\nu_{\rm pk} \sim 
\nu_s \, \gamma_p^2 \, D^2 / (1 + z) \sim 10^{19} \, \nu_{p, 19}$~Hz, 
yielding

\begin{equation}
\gamma_p \sim 10^2 \, \nu_{s,13}^{-1/2} \, D_1^{-1} \, \nu_{p, 19}^{1/2} \, .
\label{gamma_ic}
\end{equation}

Setting the observed variability timescale equal to the Compton cooling 
timescale (modulo $D/(1 + z)$) yields an estimate of the energy density 
of the external radiation field in the co-moving frame,

\begin{equation}
{u'}_s \sim 415 \, \nu_{s,13}^{1/2} \, \nu_{p, 19}^{-1/2} \, t_{{\rm var,}2}^{-1} \; 
{\rm erg \; cm}^{-3} \, , 
\label{us}
\end{equation}
which is related to the energy density in the rest frame of the host
galaxy through $u_s \approx {u'}_s / \Gamma^2 \sim {u'}_s / D^2$. Assuming that this emission originates within a few hundred 
Schwarzschild radii of
the central black hole (i.e., $R_{\rm ext} = 10^{15} \, R_{15}$~cm),
the above radiation energy density results in a luminosity of 
$L_s \sim 4 \pi \, R_{\rm ext}^2 \, c \, u_s \sim 1.6 \times 10^{42} 
\, R_{15}^2 \, D_1^{-2} \, t_{{\rm var,}2}^{-1} \, \nu_{s,13}^{1/2} \, 
\nu_{p, 19}^{-1/2}$~erg~s$^{-1}$, which corresponds to a flux of

\begin{equation}
\nu F_{\nu}^{\rm s} \sim 3.8 \times 10^{-15} \, R_{15}^2 \, D_1^{-2} \, 
t_{{\rm var,}2}^{-1} \, \nu_{s,13}^{1/2} \, \nu_{p, 19}^{-1/2} \; {\rm erg \; cm}^{-2} 
\; {\rm s}^{-1} \, ,
\label{nuFnus}
\end{equation}
which would be of the order of the observed flux of the IR peak
for $R_{15} \sim$~a few. The observed X-ray peak luminosity can be used 
analogous to Equation \ref{Lpk} to infer the density of electrons around 
$\gamma_p$:

\begin{equation}
n_p \sim 8.5 \times 10^6 \, f_{-10} \, \nu_{s,13}^{1/2} \, D_1^{-7} \, 
\nu_{p, 19}^{-1/2} \, t_{{\rm var,}2}^{-2} \; {\rm cm}^{-3} \, ,
\label{np_ic}
\end{equation}
which yields a Thomson depth of

\begin{equation}
\tau_T \sim 1.3 \times 10^{-4} \, f_{-4} \, \nu_{s,13}^{1/2} \, 
D_1^{-6} \, \nu_{p, 19}^{-1/2} \, t_{{\rm var,}2}^{-1} \, .
\label{tauT_ic}
\end{equation}
Analogous to Eqs. \ref{usy} and \ref{Lratio_KN}, we can now compute the 
expected importance of higher-order Compton scatterings through the
ratio of luminosities in second-order to first-order Compton luminosities:

\begin{equation}
\left( {L_{C2} \over L_{C1}} \right)_{\rm Thomson} \sim 2.2 f_{-10} \, 
D_1^{-8} \, \nu_{s,13}^{-1/2} \, \nu_{p, 19}^{1/2} \, 
t_{{\rm var,}2}^{-1} \, ,
\label{Lratio_ic}
\end{equation}
and this emission would peak at

\begin{equation}
\nu_{C2} \sim 10^{21} \, D_1^{-2} \, \nu_{s,13}^{-1} \, \nu_{p, 19}^2 \; {\rm Hz} \, ,
\label{nu_C2}
\end{equation}
which corresponds to $E_{C2} \sim 4 \, D_1^{-2} \, \nu_{s,13}^{-1} 
\, \nu_{p, 19}^2$~MeV and is substantially below the {\it Fermi}/LAT 
regime of $E > 100$~MeV. We therefore conclude that 
higher-order Compton scattering is not expected to lead to
a detectable signal in the {\it Fermi}/LAT or VERITAS regimes. 

\cite{disruption} suggested that the target field for Compton 
scattering might be UV -- soft X-ray emission from an accretion 
disk formed during the tidal disruption event. This would correspond 
to $\nu_{s,13} \sim 10^4$. Consequently, the X-rays could be produced
through the bulk Compton process by cold (in the co-moving frame) 
electrons. We would then infer an external radiation energy density of 
$u_s \sim 420 \, D_1^{-2} \, t_{{\rm var,}2}^{-1} \, 
\nu_{p, 19}^{-1/2} $~erg~cm$^{-3}$, corresponding to an observed 
flux of $\nu F_{\nu}^s \sim 3.8 \times 10^{-13} \, R_{15}^2 \, 
D_1^{-2} \, t_{{\rm var,}2}^{-1} \, 
\nu_{p, 19}^{-1/2}$~erg~cm$^{-2}$~s$^{-1}$. 
Considering the substantial UV extinction towards the
emission region, this flux still appears consistent with the
UVOT upper limits.

We conclude that among the scenarios discussed here, inverse-Compton 
scattering of either the observed radio -- IR radiation 
by relativistic electrons of $\gamma_p \sim 10^2$, or of a putative 
accretion-related UV radiation field scattered by cold electrons in a 
relativistically moving emission region with $D \sim 10$, are plausible 
mechanisms for the production of the observed rapidly-varying X-ray emission, and are compatible with the {\it Fermi}-LAT and VERITAS 
upper limits. As discussed by \citet{swiftpaper}, the observed SED and variability are also consistent with a synchrotron-dominated X-ray emission scenario, if the jet has a strong magnetic field (Poynting-flux-dominated) and has ongoing {\it in situ} acceleration of electrons. {However, if the energy content of the emission 
region is dominated by relativistic particles, either far 
sub-equipartition magnetic fields or an uncomfortably large
Doppler factor are required.}

\begin{acknowledgments}

{\it Acknowledgments.} This research is supported by grants from the U.S. Department of 
Energy, the U.S. National Science Foundation and the Smithsonian Institution, by NSERC in 
Canada, by Science Foundation Ireland (SFI 10/RFP/AST2748) and by STFC in the U.K.  We acknowledge the excellent 
work of the technical support staff at the Fred Lawrence Whipple Observatory and the collaborating 
institutions in the construction and operation of the instrument.  This work made use of data supplied by the UK Swift Science Data Centre at the University of Leicester.
\end{acknowledgments}

\newpage

\begin{table}[htdp]
\caption{Selection criteria used for VERITAS Analysis.  For an explanation of these parameters see Section \ref{sec:Analysis}.}
\begin{center}
\begin{tabular}{|c| c|}
\hline
Parameter & Selection Criteria \\ \hline \hline
Image size & $>$  400 digital counts ($\sim 75$ photoelectrons)\\  \hline
Mean Scaled Width & 0.05 $<$ MSW $<$ 1.15 \\  \hline
Mean Scaled Length & 0.05 $<$ MSL $<$ 1.3 \\ \hline
Height of Shower Maximum & $>$ 7 km \\ \hline
$\theta$& $<$ 0.1$^{\circ}$\\ \hline
\end{tabular}
\end{center}
\label{VERITAScuts}
\end{table}

 \begin{table}[ht]
   \begin{center}
        \caption{Data analysis results.  Flux upper limits calculated assuming a photon power-law index of -3.0 and taken at the decorrelation energy of 500 GeV.  The ratio of on-source to off-source exposure is denoted as $\alpha$.}
   \small
     \begin{tabular}{|p{60mm}||c|c|c|} \hline
	 & Total & Flaring & Low  \\ \hline \hline 
%	Date Range  & MJD 55649--55667 & MJD 55649--55651 & MJD 55652--55667 \\\hline
	Date Range  & 2011 March 29 -- April 15& 2011 March 29--31 & 2011 April 1--15 \\\hline
	ON (Source) counts  & 579 & 59 & 520 \\\hline
	OFF (Background) counts & 5639 & 604 & 5035  \\\hline
	Average $\alpha$  & 0.1 & 0.1 & 0.1 \\\hline
	Significance & $0.6 \sigma$ & $-0.3\sigma$ & $ 0.8\sigma$ \\\hline
	Excess Counts & 15.1 &  -1.4 & 16.5 \\\hline
	Flux Upper Limit ($99\%$ c.l.) [E*F(E); erg\,cm$^{-2}$\,s$^{-1}$]   & $1.4 \times 10^{-12}$ & $3.1 \times 10^{-12}$ & $1.5 \times 10^{-12}$  \\\hline
     \end{tabular}

      \label{results}
   \end{center}
 \end{table}

\end{document}